# Comparative Evaluation of Memory Technologies for Synaptic Crossbar Arrays- Part 2: Design Knobs and DNN Accuracy Trends

Jeffry Victor*, Chunguang Wang*, and Sumeet K. Gupta
Purdue University, West Lafayette, IN, USA

*Abstract*—Crossbar memory arrays have been touted as the workhorse of in-memory computing (IMC)-based acceleration of Deep Neural Networks (DNNs), but the associated hardware non-idealities limit their efficacy. To address this, cross-layer design solutions that reduce the impact of hardware non-idealities on DNN accuracy are needed. In Part 1 of this paper, we established the co-optimization strategies for various memory technologies and their crossbar arrays, and conducted a comparative technology evaluation in the context of IMC robustness. In this part, we analyze various design knobs such as array size and bit-slice (number of bits per device) and their impact on the performance of 8T SRAM, ferroelectric transistor (FeFET), Resistive RAM (ReRAM) and spin-orbit-torque magnetic RAM (SOT-MRAM) in the context of inference accuracy at 7nm technology node. Further, we study the effect of circuit design solutions such as Partial Wordline Activation (PWA) and custom ADC reference levels that reduce the hardware non-idealities and comparatively analyze the response of each technology to such accuracy enhancing techniques. Our results on ResNet-20 (with CIFAR-10) show that PWA increases accuracy by up to 32.56% while custom ADC reference levels yield up to 31.62% accuracy enhancement. We observe that compared to the other technologies, FeFET, by virtue of its small layout height and high distinguishability of its memory states, is best suited for large arrays. For higher bit-slices and a more complex dataset (ResNet-50 with Cifar-100) we found that ReRAM matches the performance of FeFET.

## I. INTRODUCTION

THE success of Deep Neural Networks (DNNs) for large number of tasks such as computer vision, autonomous vehicles, etc., has increased the demand for their specialized and efficient hardware [1]. Unfortunately, von-Neumann architectures such as CPUs and GPUs are limited by the memory-processor data transfers and are not suitable to meet the computation demands of DNNs [1]. Hence, in-memory computing (IMC), which alleviates the von-Neumann bottleneck, has been explored as an alternative to standard platforms for DNN acceleration.

The common IMC solution to accelerate DNNs is to employ a crossbar memory array to efficiently implement Matrix- Vector Multiplication (MVM), which is the most frequent and data-intensive operation in DNNs [1]. This avoids the expensive data-movement as the MVM computation occurs within the memory macro. However, multiple *cross-layer challenges* limit the pervasiveness of this approach.

At the device level, the inherent non-linearities and variability in memory technologies negatively impacts the robustness of MVM computation in crossbar arrays [2]. At the circuit level (as shown in Fig. 1), parasitic resistances (such as wire and driver resistances) induce IR-drops, leading to data-dependent deviations of the actual output from the expected output causing IMC errors [2]. Furthermore, such non-idealities are aggravated in ultra-scaled nodes due to higher wire resistance and increased device non-idealities (such as leakage) [3], thus restricting IMC from reaping the expected benefits of technology scaling. To make the matters more complex, the design knobs that enhance the performance/density metrics (such as increasing in the number of bits per device or enhancing the parallelism of IMC) sometimes are in conflict with the IMC robustness [2].

Therefore, mitigating the hardware non-idealities has been explored in earnest to enhance inference accuracy and to expand the design space of DNN accelerators [4], [5], [6], [7],

This work was supported in part by the Center for Brain-Inspired Computing (C-BRIC), and in part by the Center for the Co-Design of Cognitive Systems (COCOSYS), funded by Semiconductor Research Corporation (SRC) and DARPA under Grant AWD-004311-S4. (Corresponding author: Jeffry Victor.)

Jeffry Victor, Chunguang Wang, and Sumeet K. Gupta are with the Elmore Family School of Electrical and Computer Engineering, Purdue University, West Lafayette, IN 47907 USA (e-mail: louis8@purdue.edu; wang4015@purdue.edu; guptask@purdue.edu).

* These authors contributed equally to this paper.

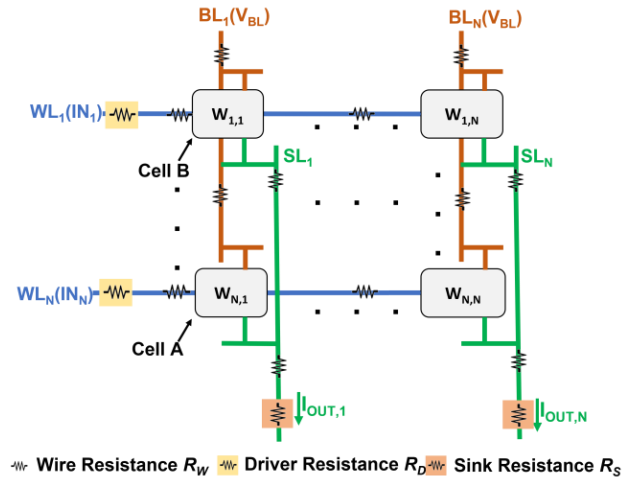

Fig. 1 Crossbar array with wire-resistances and other parasitic resistances.

[8], [9]. These solutions include hardware-aware training [6], re-mapping weights [7], [10], partial wordline activation (PWA) [9] and others.

Though such techniques have been shown to be effective in mitigating the non-idealities in crossbar arrays and improve DNN inference accuracy, there is a lack of understanding of how different memory technologies and their unique attributes interact with such design approaches and what the relative advantages and disadvantages of various technologies are when such techniques are applied. Furthermore, the analysis of these techniques has been mainly conducted in relatively older technologies, in which the hardware non-idealities are fairly manageable. Therefore, the efficacy of these techniques in deeply scaled technologies needs a systematic investigation.

To gain insights into the device-circuit interactions in crossbar arrays in deeply scaled technologies that govern the IMC robustness, this paper builds upon our analysis in Part 1 and conducts a comparative technology evaluation at the 7nm node by analyzing the response of various memory technologies to the design knobs used in crossbar array design. The key contributions of this work are as follows:

- We comparatively evaluate 8T SRAMs, ferroelectric transistors (FeFETs), spin orbit torque magnetic memories (SOT-MRAMs) and resistive memories (ReRAMs) and analyze the effect of cross-layer design knobs such as increasing array size, increasing bit slice, and changing datasets/networks on crossbar non-idealities and inference accuracy.
- We study the response of the four memory technologies to PWA in terms of its efficacy of improving DNN accuracy.
- Based on an extensive analysis of the IMC sense margins, we propose replacing the standard linear references of analog-to-digital converters (ADCs) with hardware non-ideality-aware reference levels and evaluate its efficacy in boosting the inference accuracy for different technologies.

## II. BACKGROUND

### A. Related Works

The implementation of MVM on an IMC platform typically involves (i) decomposing the weight matrix into smaller sub-matrices for deployment on crossbar arrays of practical sizes, (ii) bit-slicing the weights and encoding the bits as binary or multi-level conductance of bit-cells, (iii) encoding the input activations as word-line voltage bit streams, (iv) sensing the analog output current or voltage on the sense-line (SL) (v) converting it into a digital partial sum using an ADC and (vi) combining all the partial sums from different bit-slices and accumulating the partial sums from different sub-matrices in the digital domain to obtain the final MVM product.

One of the main challenges in such a design is the presence of crossbar non-idealities such as parasitic resistances (wire resistance, source/sink resistance), and device non-linearities which lead to a deviation of the SL current from its ideal value impairing DNN accuracy. [NO_PRINTED_FORM]. Further,

TABLE I
Inference Accuracy of ResNet-20 on Cifar-10 on 64x64 crossbar arrays across technologies

| Technology | FWA / PWA | $R_{HRS}/R_{ON}$ | $R_{OFF}/R_{ON}$ | Cell Height (GP) | Accuracy (%) |
|---|---|---|---|---|---|
| SRAM | FWA | $3.5*10^6$ | $2.5*10^6$ | 2 | 88.8 |
| FeFET | FWA | 66 | $3.8*10^2$ | 1 | 91.6 |
| RERAM | FWA | 38.3 | $3.5*10^3$ | 1.5 | 91.0 |
| SOT-MRAM PWA 8 | PWA-8 | 5 | $1.0*10^4$ | 2 | 79.6 |

GP – Gate Pitch

cross-layer design knobs such as array size, bit slice and synaptic technology type affect such non-idealities and consequently impact the IMC robustness. For instance, large crossbar arrays offer more parallelism and higher array efficiency but suffer from increased IR drops [11]. In this context, Sharma *et al.* [11] have analyzed non-idealities in SOT-MTJ crossbars at 65 nm technology node for different array sizes.

Bit slice is another important design knob which enhances memory footprint and parallelism, and reduces peripheral circuit complexity. However, it worsens the IR drop in the crossbar array [12]. To understand the effect of bit slice on DNN inference accuracy, Chakraborty *et al.* analyzed the inference accuracy on ReRAM crossbars at 65nm for different bit slices [2]. They observed a degradation in accuracy with an increase in bit slice beyond 2.

ADC precision is another critical factor which determines the energy consumption and sensing robustness of an IMC macro [13], [14]. Large arrays require high ADC precisions, which incurs huge energy and area overheads [13]. Some works have explored PWA as a solution to this problem [9]. By asserting only a subset of all WLs, the ADC precision and its overheads can be reduced. Yi *et al.* have also explored the use of PWA to increase inference accuracy in ReRAM crossbars [9] as PWA limits the output current and helps in managing the IR drops. However, it is important to note that PWA reduces parallelism when compared to asserting all WLs concurrently (referred to as Full Wordline Activation (FWA) in this paper).

Though the above-mentioned works provide many useful insights into the cross-layered nature of these design knobs, there are still two important gaps. First, these works have analyzed the performance of only one memory technology in their respective works. Hence, a comparative analysis of the behavior of different memory technologies in response to these design knobs is lacking. Second, most of these works are focused on somewhat older technology nodes (such as 65 nm in [13]) in which the hardware non-idealities are fairly manageable. The question is how the crossbar design knobs and non-idealities interact in a deeply scaled technology. To address both these gaps, in this paper, we *comparatively analyze the behavior of different memory technologies to the crossbar array design knobs at 7 nm technology node*.

### B. Recap of Part I

In Part I of this paper [15], we had performed a comparative evaluation of four memory technologies in the context of IMC robustness and inference accuracy at 7 nm technology node: 8T

SRAM, FeFET, ReRAM and SOT-MRAM. We included the device-circuit non-idealities utilizing the Genie-X framework [2] appropriately modified for our analysis. We analyzed device-circuit optimization knobs to obtain optimal design points for each technology for a fair comparative technology evaluation. The design parameters were chosen considering non-ideality factor (NF) and sense margin (SM). The NF quantifies the deviation of the non-ideal current ($I_{non-ideal}$) from the ideal value ($I_{ideal}$) due to hardware non-idealities [2]. It is defined as:

$$NF = \left| \frac{I_{ideal} - I_{non-ideal}}{I_{ideal}} \right| \quad (1)$$

Technologies and array designs with smaller NFs are less impacted by non-idealities and are likely to exhibit higher DNN accuracy. In addition to NF, SM is another important metric to consider to account for the ADC limitations. SM between output x and output x-1 is defined as:

$$SM_X = \frac{I_{X,MIN} - I_{X-1,MAX}}{2}$$

Here, $I_{X,MIN}$ ($I_{X-1,MAX}$) is the minimum (maximum) non-ideal current when the output is x (x-1). We obtained these currents from extensive SPICE simulations based on DNN workloads. A high SM implies high sensing robustness whereas a negative SM may lead to most currents being assigned a wrong output state, leading to computation errors.

We used the NF and SM to make design choices for 64x64 crossbar arrays. As an example, we observed that applying input on the gate of the access transistor/FeFET (G-input) vis-à-vis on one of the current-carrying terminals (D-input) helps in mitigating the effect of non-idealities. We also included a dummy column in each crossbar to minimize the effect of high resistance state (HRS) currents on IMC robustness. Further, we optimized crucial parameters such as ON resistance of the bit-cells ($R_{ON}$ corresponding to input =1 and weight=1). Besides such technology-agnostic optimizations, we analyzed technology-specific design knobs such as ferroelectric thickness optimization for FeFETs and co-optimization of tunneling oxide thickness and PWA for SOT-MRAMs. Based on such optimal design choices, we compared the DNN inference accuracy for each technology by implementing matrix-vector-multiplication (MVM) operations on the optimized crossbars based on each technology.

We observed the non-ideal inference accuracy of the four technologies in 64x64 arrays for ResNet-20 and CIFAR-10 dataset. We found the FeFETs exhibit the highest inference accuracy followed closely by ReRAM (Table I). SRAM exhibit a moderately lower accuracy than ReRAM. SOT-MRAM display the least inference accuracy amongst the four technologies and needed PWA (of 8 WLs per cycle) to obtain acceptable levels of accuracy. Note from Table I, the other three technologies employ FWA for maximum parallelism.

We explained these trends (in Part 1) using the characteristics of these technologies, including (i) layout height of the bit-cell (which dictates the wire resistance), (ii) resistances corresponding to different input($In$)-weight($W$) combinations ($R_{ON}$: $In=1,W=1$, $R_{HRS}$: $In=1,W=0$; $R_{OFF}$: $In=0,W=1$ and $R_{OFFH}$: $In=0,W=0$) and (iii) the sensitivity of the bit-cell current to the IR-drops.

FeFETs exhibit the lowest layout height (least wire resistance) and a reasonably high $R_{HRS/OFF}$:$R_{ON}$, which ensures a high inference accuracy of 91.6% (only ~1% less than the software (ideal) baseline).

ReRAM has a taller bit-cell than FeFET and so is more impacted by wire-resistance. However, the 1T-1R configuration of ReRAM, which is typically designed for the access transistor to have a much lower resistance that the memory device, helps in reducing the sensitivity of the bit-cell current to IR drops. Specifically, IR drops on the SL leads to source-degeneration of the transistor it is connected to (for all technologies). The 1T-1R configuration of ReRAM reduces the impact of source-degeneration on the bit-cell current as the dominant resistance comes from the memory device. This compensates to some extent for the detrimental effects of a taller bit-cell. (Note, in FeFETs and SRAMs, source-degeneration has a strong effect on the bit-cell current). Additionally, $R_{HRS,OFF}$:$R_{ON}$ of ReRAM is similar to that of FeFETs. Thus, the 1T-1R configuration and the high $R_{HRS/OFF}$:$R_{ON}$ enable ReRAM to achieve an inference accuracy of 91%, only marginally smaller than FeFETs.

The read port of 8T-SRAM cell features two series connected transistors that make the bit-cell current significantly susceptible to source degeneration due to IR drops. This issue is exacerbated by the larger cell height of SRAM (compared to FeFET and ReRAM), which results in high wire resistance. However, SRAM has significantly high $R_{HRS/OFF}$:$R_{ON}$. Overall, we observed a lower inference accuracy of 88.8% (compared to FeFET and ReRAM).

Though SOT-MRAM benefits from a 1T-1R cell configuration similar to ReRAM, its cell height is as tall as that of SRAM. Additionally, SOT-MRAM has a very poor $R_{HRS}/R_{ON}$. As a result, SOT-MRAM requires PWA of 8 to achieve a comparable accuracy of 79.6%, which is still significantly lower than that of the other technologies.

Table 1 summarizes the key findings of Part 1. For a deeper analysis of the relationship between cell height, $R_{HRS}/R_{ON}$, $R_{OFF}/R_{ON}$ and accuracy, as well as the parameters for each technology, please refer to Part 1. In this part, we analyze how 8T SRAM, FeFET, ReRAM and SOT-MRAM (optimized based on Part I) respond to various design knobs such as array size, bit-slice, change in dataset/architecture and non-ideality mitigation techniques and comparatively evaluate the technologies in the context of these design options.

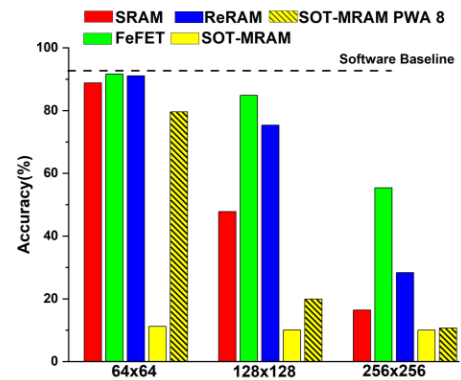

Fig. 2. Effect of array size increase on accuracy for each technology.

### III. EFFECT OF CROSSBAR ARRAY SIZE

In this section, we analyze the effect of array size on inference accuracy for the four technologies and compared their response. As shown in Fig. 2, all the technologies exhibit a degradation in accuracy with an increase in array size due to longer wires (higher wire resistance) and larger number of weight bits per column (that increases output range). Comparing the technologies, we observe that FeFETs maintain the highest accuracy across all array sizes. FeFETs also demonstrate greater resilience to increase in array size (i.e. lower sensitivity of accuracy to array size) compared to other technologies. This can be attributed to the smallest layout height of FeFETs, which reduce the effect of increasing wire resistance with array size (compared to the other technologies). When the array size increases from 64 to 256, the accuracy for FeFETs decreases from 91.6% to 84.85%.

ReRAM exhibits inference accuracy comparable to that of FeFET for 64x64 array. However, for 128x128 and 256x256 arrays, its accuracy is 10.5% and 27.54% lower than that of FeFET, respectively. Despite its advantageous 1T-1R configuration (i.e. low sensitivity of bit-cell current to IR drops) and a similar $R_{HRS/OFF}:R_{ON}$ to FeFET, the impact of increasing array size aggravates due to larger layout height (and larger wire resistance) of ReRAM than FeFET.

SRAM exhibits sharper decline in accuracies compared to ReRAM and FeFET (accuracy = 88.8%, 55.3% and 47.7% for the 64x64, 128x128, and 256x256 arrays, respectively). This is mainly due to the largest layout height of SRAM, which significantly increases the wire resistance.

SOT-MRAM exhibits inference accuracy of ~10% across various array sizes, when FWA is used. This consistently low accuracy is due to the combined effects of its tall cell height (high IR drops), and its inherently low distinguishability ($R_{HRS}/R_{ON}$). To get around this issue, we use PWA of 8 WLs to enhance the accuracy for SOT MRAM to 79.6% for a 64x64 array. However, as the array size increases to 128x128 and 256x256, inference accuracy drops to unacceptably low 19.9% and 10.7%, respectively. This suggests the array size for SOT-MRAMs, despite PWA, is limited to 64x64. It may be mentioned that if the sense margin constraints can be relaxed (i.e. with better ADC designs) to lower the SL current, it may be possible to enhance the accuracy for larger array sizes (which needs further analysis; here, we restrict our analysis to designs in Part I based on maximizing the sense margins). However, we expect that achieving respectable accuracy levels may still be challenging, mainly due to their low $R_{HRS}/R_{ON}$ and high wire resistance at 7nm node.

### IV. NON-IDEALITIES-MITIGATION TECHNIQUES

To mitigate the effect of increasing array size on the inference accuracy, we explore two techniques that can help in reducing the non-idealities. First, we analyze the PWA for all the technologies which reduces the magnitude of SL current and hence, the IR drop. Second, we explore a technique which improves the SM by using ADCs with custom reference levels. In this work, we focus on 128x128 arrays with an aim to bring their accuracy as close to the software baseline as possible. For 256x256 arrays, we anticipate that due to longer wire lengths and higher number of output states per column, these techniques will be less effective in closing the gap between the non-ideal and software accuracies.

#### A. Partial Wordline Activation (PWA)

FWA is typically utilized to achieve the full benefits of parallelism in crossbar enabled MVM operations. However, this implies large SL currents and high IR drops (especially for large array sizes). To mitigate the non-idealities, PWA can be utilized to assert only a subset of all WLs in one cycle, thereby reducing SL current and the IR drop, albeit at the cost of reduced parallelism.

Recall from Fig. 2, the 64x64 arrays of SRAM, FeFET and ReRAM have an inference accuracy reasonably close to the software baseline. Hence, we choose to perform PWA of 64 WLs on 128x128 crossbars of these three technologies, thus taking two cycles to complete one MVM operation. Also, recall, SOT-MRAMs with PWA of 8 yield unacceptably low accuracies for 128x128 arrays (Fig. 2). Therefore, we choose to perform PWA of 4 WLs per cycle for a 128x128 SOT-MRAM crossbar, thus taking 32 cycles to complete one MVM operation. Note that, further reducing the number of WLs asserted per cycle increases inference accuracy mildly but significantly increases the MVM latency.

The results of our analyses are shown in Fig. 3. We observe that with PWA of 64 WLs, FeFET and ReRAM exhibit accuracy which is reasonably close to the software baseline. It is interesting to note in Fig. 3 that the increase in accuracy due to PWA is larger for ReRAM than FeFET, (which leads to comparable accuracies for the two technologies). This can be explained as follows. For the top 64 cells in a column of the 128x128 array (which are farthest from the ADC), the wire resistance associated with SL running through the bottom half of the array leads to large source degeneration. Although PWA reduces the source degeneration (compared to FWA due to lower SL current), this effect is still significant. Now recall, the major non-ideality-inducing component in FeFETs is the source degeneration (as other factors such as wire resistance are small due to compact layout). On the other hand, ReRAM suffer mainly from high wire resistance (due to less compact layout) while the effect of source degeneration is highly reduced due to 1T-1R configuration (see Section II). Therefore, when we apply PWA, the major non-ideality component in ReRAM is targeted, which along with insignificant effect of source degeneration due to 1T-1R bit-cell contributes to a large increase in the accuracy. However, for FeFETs, the prime non-ideality (source degeneration effect) is reduced but is still significant in the top half of the column. Thus, PWA increases the accuracy to a larger extent in ReRAM.

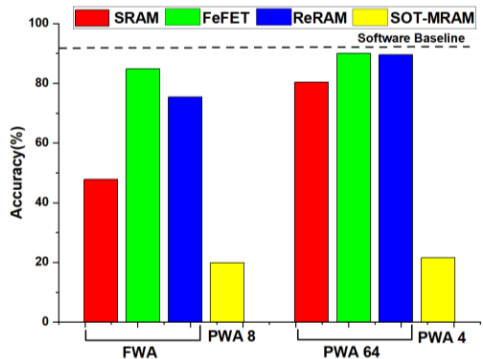

Fig. 3. Effect of PWA on accuracy for each technology at 128x128 array size.

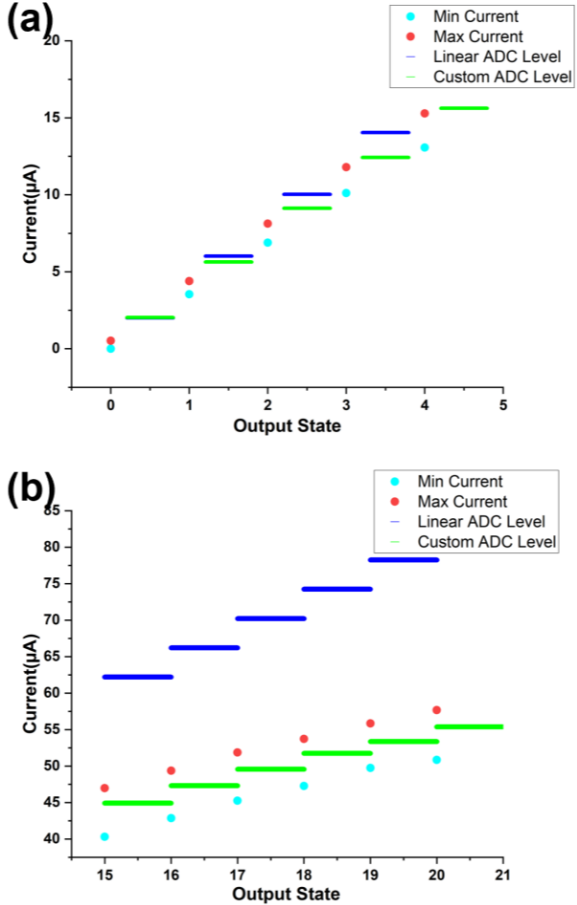

Fig. 4. Plots showing linear and custom ADC levels for outputs with (a) positive SM and (b) negative SM

For SRAMs, PWA improves the accuracy to 80.34%. Since FWA in SRAM suffers from large IR drops due to its tall cell height, the application of PWA presents the largest scope for improvement in accuracy as it targets the major non-ideality component. Note, this improvement is partially offset by the source degeneration effect in SRAMs, but the accuracy increase is still quite high.

From Fig. 3, we also observe that for SOT-MRAM, PWA of 4 WLs increases the inference accuracy to 21.6% from 19.9% (PWA of 8). Thus, even with PWA of 4, the inference accuracy cannot be gained appreciably for a 128x128 array. This reinforces our conclusion that SOT MRAMs are limited to 64x64 array (with a PWA of 8) at 7nm node due to their low TMR and high wire resistance.

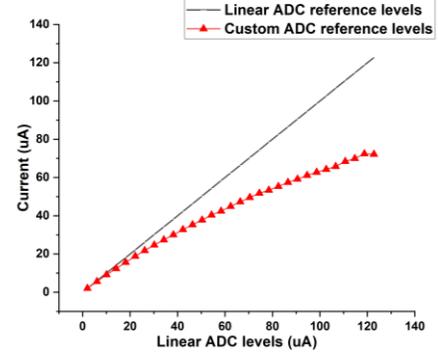

Fig. 5. Plot showing the deviation of custom ADC levels from Linear ADC levels

*B. Custom ADC reference levels*

Typically, crossbar-enabled MVM operations utilize linearly separated ADC reference levels to convert the SL current into digital values. These linear ADC reference levels are based on ideal SL currents which provide constant and positive sense margin (SM) between any two consecutive output states. But the presence of non-idealities introduce non-linearity in the SL currents with respect to the expected output value. Hence, the linear ADC reference levels sub-optimal for sensing non-ideal SL currents. Hence, in this section, we explore the use of custom ADC reference levels based on non-ideal currents for each output state and analyze the response of each technology to this technique.

We choose a custom ADC reference level to be at the average of maximum current of state x-1 and minimum current of state x. This implies that SL currents lower (higher) than a custom ADC reference level are assigned to the state x-1 (x). We obtain the minimum and maximum current of each state from extensive SPICE simulations of the four technologies (for more details refer to Part 1). To observe the difference in sensing robustness between linear and custom ADC reference levels, we plot the minimum and maximum current for each output state along with linear and custom ADC reference levels in Fig. 4. This data corresponds to FeFET-based 128x128 array. However, the technique discussed here is generally applicable.

For low magnitude output states (Fig. 4a), we observe that the SM is positive, but decreases as the output increases. The linear and custom ADC reference levels are located within the range of the output currents but the difference between them increases as the output increases. This is due to the linear ADC reference levels being chosen based on ideal SL currents which do not account for IR-drop-induced current reduction.

When we analyze the high magnitude outputs (Fig. 4b), we observe a more significant impact of the hardware non-idealities due to larger SL current and higher IR drops. Even for FeFETs – the best performing technology in our analysis, we observe negative SM for high output states > 15. As a result, there is an overlap of current ranges between adjacent states.

The trends in Fig. 4b also indicate that the linear ADC reference levels are outside the range of corresponding currents by a large amount. This results in significant sensing errors. On the other hand, the custom ADC reference levels mitigate this issue. However, even then, for high outputs, the custom ADC reference levels are not always able to correctly assign output states to SL currents. But it is important to note that the error is at most +/-1, which is significantly less compared to the linear ADC reference levels. Fig. 5 illustrates the increasing divergence between linear and custom ADC reference levels, suggesting that this technique targets errors in higher output states. It is most effective when the SM is positive, but it mitigates the effect of negative SM on the magnitude of the error.

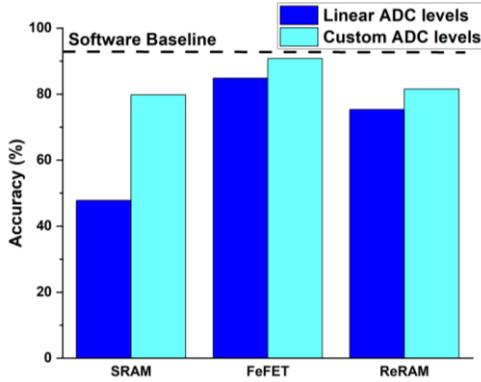

Fig. 6. Inference accuracy using Linear and Custom ADC levels

Note that, due to the highly sparse nature of DNN workloads, the low magnitude output states are more frequent than high magnitude output states. Hence, even the linear ADC reference levels sense a significant number of SL currents correctly. Most of the errors occur for large output states and the error magnitude increases with the output. The custom ADC reference levels are able to sense a larger range of outputs correctly and reduce the magnitude of the error for higher outputs.

In Fig. 6, we analyze the effect of this technique on DNN inference accuracy. Custom ADC reference levels improve the inference accuracy of FeFET from 84.85% to 90.8%. FeFET experiences lesser IR drop-induced current reduction compared to other technologies due to its small cell-height, resulting in low sensing errors, even with linear ADC reference levels. Custom ADC reference levels further reduce sensing errors, bringing the inference accuracy closer to the software baseline.

For ReRAM, custom ADC reference levels improve inference accuracy from 75.3% to 81.5%. Due to a larger cell height compared to FeFET, ReRAM experiences higher IR drop-induced current reduction. Consequently, its SM is negative for a larger number of output states compared to FeFET. As a result, while custom ADC reference levels enhance the accuracy, they are less effective for ReRAM than FeFET.

On the other hand, SRAM exhibits a significant increase of 31.62% in inference accuracy (from 47.7% to 79.4%) with custom ADC reference levels. While the tall SRAM cell leads to large IR drop-induced non-linearity in SL currents, the range of currents for each output is reduces due to its excellent $R_{HRS/OFF}:R_{ON}$. In other words, while the deviation of the non-ideal current from the ideal value is high, SM (determined mainly by the range of currents for an output) is positive for many states. As custom ADC reference levels track the deviation and are most effect for positive SMs, a large inference accuracy improvement of 31.62% is achieved.

For SOT-MRAM (not shown in Fig. 6), we observe that the sense margins are negative for a large range of outputs due to their poor $R_{HRS}/R_{ON}$ ratio. Thus, custom ADC reference levels do not make any significant enhancement to the inference accuracy of SOT MRAM.

To qualitatively estimate the cost of the aforementioned custom ADC reference level solution, we refer to the nonlinear successive approximation register (SAR) ADC proposed in [16]. This ADC implements power-of-two quantization by pre-charging the reference capacitor to select voltage levels [16]. The custom ADC design can be implemented by following a similar approach, wherein the pre-charging of the reference capacitor is adjusted to a predetermined level, rather than conforming to the linear levels typically used in a standard SAR ADC.

*C. Comparison of PWA and custom ADC reference levels*

Let us now compare the efficacy of PWA and custom ADC reference levels for different technologies. Our results show that PWA is more effective in increasing the inference accuracy compared to the custom ADC reference levels. However, these benefits come at the cost of reduced parallelism. In contrast, the custom ADC reference levels achieve full parallelism, but require some modifications to the ADC circuit design, as discussed previously. It is interesting to note that PWA not only leads to reduced non-idealities but also the lowers the range of outputs. While SM of the low magnitude output states is high even in FWA (as discussed in the previous sections), lower non-idealities in PWA further enhance their SM. Thus, PWA can potentially by used synergistically with custom ADC reference levels; however, the analysis of the implications of this combination requires further investigation (which is beyond the scope of this work).

Scaling these techniques to larger array sizes should involve careful consideration of multiple factors. In our analysis, we have used a PWA of 64 for 128x128 arrays. But larger arrays may require PWA of lesser number of WLs to mitigate the increase in wire-length. This will increase the number of cycles needed for MVM, leading to larger throughput penalties with PWA. Another effect of increasing array size is increase in the source degeneration effect, which reduces the magnitude of SL current. This causes the linear ADC reference levels to perform poorer for larger arrays. Hence, custom ADC reference levels become more lucrative for larger arrays. However, one must keep in mind that SM becomes increasingly negative with array size. Hence, custom ADC reference levels can become less effective beyond a certain array size.

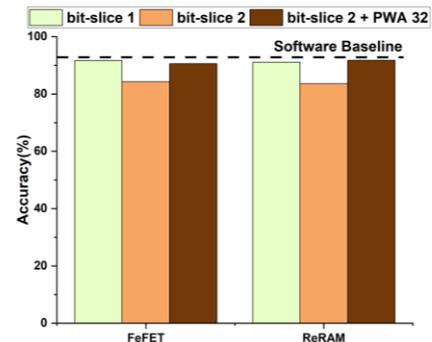

Fig. 7. Effect of varying bit-slice and PWA on accuracy of FeFET and RERAM

## IV. INCREASING THE BIT SLICE

In this section, we analyze the effect of increasing bit slice (number of bits per device) on the accuracy for ReRAM and FeFET (the technologies capable of storing > 2 levels per device). Increasing the bit slice reduces the number of crossbars required albeit increases the precision (and energy) of ADCs as the number of output states increase. Using higher bit-slices can be particularly attractive for highly area-constrained platforms, which cannot fit the entire neural network model. For such designs, fetching the weights from dynamic RAMs (DRAMs) becomes necessary incurring energy and latency costs. In this context, increasing the storage capability of crossbars via multi-level storage can be useful to lessen the DRAM access costs. However, increasing the bit-slice also reduces the inference accuracy due to higher non-idealities, as shown previously in [2]. Here, we comparatively analyze how FeFET and ReRAM respond to increasing the bit-slice. Since the analysis in [2] has suggested keeping the bit-slice to be at or below 2, we focus on analyzing the performance of FeFET and ReRAM considering 1 or 2 bits stored in each device. In this section, we focus on 64x64 arrays.

In our analysis, the conductance for weight=0 and weight=1 is the same in bit-slice of 1 and 2. This is to maintain sufficient sense margin, even for bit-slice of 2. This implies that the conductance for weight=2 (3) need to be chosen to be about 2 (3) times that of weight=1 to preserve the distinguishability between states. This high conductance leads to an increase in the magnitude of SL current, which reduces inference accuracy for bit-slice 2 in ReRAM and FeFET compared to bit-slice of 1 (Fig. 7).

The inference accuracy for FeFET decreases from 91.6% to 83.5%, while that for ReRAM decreases from 91% to 84.2% when transitioning from bit-slice of 1 to bit-slice of 2. While FeFETs exhibit the advantage of compact bit-cell (smaller wire resistance), ReRAM show lower sensitivity of bit-current to IR drops (due to its 1T-1R configuration). These two factors counterbalance to yield similar drop in accuracy for both the technologies.

To mitigate this accuracy drop, we employ PWA of 32 WLs in both ReRAM and FeFET arrays. This increases accuracy close to bit-slice of 1. With PWA, the accuracy of FeFET in bit-slice 2 improves to 91.7%, while that of ReRAM reaches 90.5%.

## V. CHANGING NETWORK AND DATASET

To compare the performance the four technologies for a more complex dataset and network, we analyze Resnet (RN)-50 on CIFAR (CF)-100 considering 64x64 array, bit-slice 1 and linear ADC reference levels for all technologies [17], [18]. Note that, we have utilized SOT-MRAM with PWA of 8 WLs given its higher susceptibility to non-idealities, while for the other three technologies, we utilize FWA.

The results are illustrated in Fig. 8. For RN-50 + CF-100, the software baseline accuracy is 72.8%. The inference accuracies for FeFET, ReRAM, SRAM, and SOT-MRAM (with PWA of 8) are 71.1%, 70.4%, 67.8%, and 62.4%, respectively. FeFET and ReRAM exhibit comparable accuracy, followed by SRAM and SOT-MRAM. Recall, in the case of RN-20 + CF-10, the software baseline accuracy is 92.8%. The corresponding inference accuracies are 91.6%, 91%, 88.8%, and 79.6%, respectively (also shown in Table I). The accuracy drop relative to the software baseline is higher in RN-50 + CF-100 than in RN-20 + CF-10 for FeFET, ReRAM, and SRAM. However, for SOT-MRAM with PWA 8, the accuracy drop is higher in RN-20 + CF-10. This can be attributed to the interaction between two attributes of the networks: (1) sparsity and (2) sensitivity to computational errors.

RN-50 exhibits higher sparsity compared to RN-20. This reduces the output current range in crossbar arrays, thereby lowering sensing errors and enhancing accuracy. However, the deeper architecture of RN-50 allows errors to accumulate across layers, increasing the network's sensitivity to errors which can potentially reduce accuracy.

For SOT-MRAM, the use of PWA limits the output range and the non-idealities. Further, for RN-50, due to its inherently higher sparsity than RN-20, the range of outputs is even lower. Since SOT MRAMs offer robust computations only for a small output range (due to their low TMR), RN-20 is affected more by the non-idealities, while RN-50 benefits from the combined effect of high sparsity and PWA. This reduction in computational errors leads to lower accuracy drop in RN-50+CF-100 (despite its inherently higher sensitivity to errors). On the other hand, for FeFETs, ReRAM, and SRAM, the impact of hardware non-idealities is much lower than SOT MRAMs. Thus, while higher sparsity in RN-50 is beneficial, its role in reducing the computational errors is less due to the inherent robustness of the three technologies. In this case, the network sensitivity to computational errors plays a more significant part, leading to lower accuracy drop in RN-20+CIFAR-10 than RN-50+CIFAR-100.

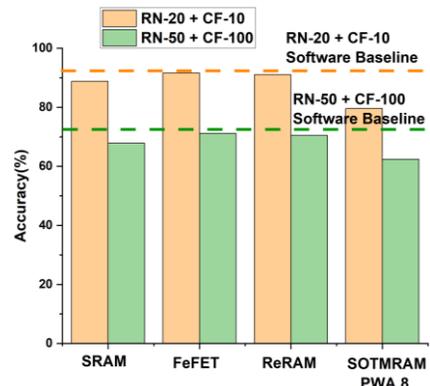

Fig. 8. Inference Accuracy results of Resnet-50 on Cifar-100 and ResNet-20 on Cifar-10.

## VI. Conclusion

This work comparatively evaluated four memory technologies – 8T SRAM, FeFET, ReRAM and SOT-MRAM - at a deeply scaled (7 nm) node in the context of crossbar-based IMC robustness and inference accuracy for RN-20 on CF-10 dataset. We analyzed the impact of various design knobs such as array size and bit slice on the performance of these technologies. Additionally, we evaluated two non-idealities-mitigation techniques - PWA and custom ADC reference levels - and the response of each technology to these design solutions. Our analysis shows that amongst the four technologies, FeFETs, by virtue of small layout height and high $R_{HRS/OFF}:R_{ON}$ ratios, are the most suitable for larger array sizes, exhibiting the highest accuracy and least accuracy degradation with increasing array size. Our analysis also indicates that PWA improves DNN inference accuracy significantly and the accuracy enhancement due to PWA is higher in ReRAM than FeFETs, leading to comparable accuracies in these two technologies (and higher than SRAMs and SOT-MRAMs). Accuracy improvements due to custom ADC reference levels are also significant but marginally lower than PWA. Also, these benefits do not incur expensive latency overheads, unlike PWA. Furthermore, for bit-slice of 2, we observed that ReRAM matches the performance of FeFET. For a more complex dataset (RN-50 with CF-100), ReRAM and FeFET exhibit comparable accuracy, followed by SRAM and SOT-MRAM (with PWA of 8). We discussed how the response of the technologies to changing dataset and DNN architecture is dictated by an interplay of multiple factors such as sparsity, network sensitivity to errors and the inherent robustness of a technology to crossbar non-idealities.

### Acknowledgement

We would like to thank X. Chen, A. Malhotra, K. Cho, T, Sharma and K. Roy at Purdue University for various useful discussions.